\documentclass[11pt]{article}

\usepackage{amsmath,amssymb,latexsym}

\catcode`\@=11  %%%% so that we can use LaTeX's private commands 
\def\newrmtheorem#1{\@ifnextchar[{\@ormthm{#1}}{\@nrmthm{#1}}}

\def\@nrmthm#1#2{%
\@ifnextchar[{\@xnrmthm{#1}{#2}}{\@ynrmthm{#1}{#2}}}

\def\@xnrmthm#1#2[#3]{\expandafter\@ifdefinable\csname #1\endcsname
{\@definecounter{#1}\@addtoreset{#1}{#3}%
\expandafter\xdef\csname the#1\endcsname{\expandafter\noexpand
  \csname the#3\endcsname \@thmcountersep \@thmcounter{#1}}%
\global\@namedef{#1}{\@rmthm{#1}{#2}}\global\@namedef{end#1}{\@endtheorem}}}

\def\@ynrmthm#1#2{\expandafter\@ifdefinable\csname #1\endcsname
{\@definecounter{#1}%
\expandafter\xdef\csname the#1\endcsname{\@thmcounter{#1}}%
\global\@namedef{#1}{\@rmthm{#1}{#2}}\global\@namedef{end#1}{\@endtheorem}}}

\def\@ormthm#1[#2]#3{\expandafter\@ifdefinable\csname #1\endcsname
  {\global\@namedef{the#1}{\@nameuse{the#2}}%
\global\@namedef{#1}{\@rmthm{#2}{#3}}%
\global\@namedef{end#1}{\@endtheorem}}}

\def\@rmthm#1#2{\refstepcounter
    {#1}\@ifnextchar[{\@yrmthm{#1}{#2}}{\@xrmthm{#1}{#2}}}

\def\@xrmthm#1#2{\@beginrmtheorem{#2}{\csname the#1\endcsname}\ignorespaces}
\def\@yrmthm#1#2[#3]{\@opargbeginrmtheorem{#2}{\csname
       the#1\endcsname}{#3}\ignorespaces}

\def\@beginrmtheorem#1#2{\trivlist \item[\hskip \labelsep{\bf #1\ #2}]}
\def\@opargbeginrmtheorem#1#2#3{\trivlist
      \item[\hskip \labelsep{\bf #1\ #2\ (#3)}]}

\catcode`\@=12  %%%%  back to non-character 

\newtheorem{thm}{\bf Theorem}[section]
\newtheorem{lem}[thm]{\bf Lemma}        %% lemmas, props, cor, etc
\newtheorem{prop}[thm]{\bf Proposition}  %% with the theorems.
\newtheorem{cor}[thm]{\bf Corollary}
\newrmtheorem{deff}[thm]{\bf Definition}

\newrmtheorem{remark}[thm]{\bf Remark}

\newcommand{\Bbox}{
{\unskip\nobreak\hfil\penalty50
\hskip1em\hbox{}\nobreak\hfil{\lower .5pt \hbox{$\Box$}}
\parfillskip=0pt \finalhyphendemerits=0 \par}
}

\newcommand{\eop}{
\ifmmode {\hbox{\Bbox}} \else \Bbox \fi
}

\def\os{{\,\times\,}}

\def\to{{\,\rightarrow\,}}

\def\1#1{
\hbox{{\bf 1}$_{#1}$}
}

\def\C{{\mathcal{C}}}
\def\bQ{{\mathbf{Q}}}
\def\cQ{{\mathcal{Q}}}
\def\cG{{\mathcal{G}}}
\def\x{{\times}}
\def\le{{\langle}}
\def\re{{\rangle}}
\def\dr{{^\dagger}}
\def\Cpo{{\bf Cpo}}

\usepackage{amsmath, amssymb}

\begin{document}

\title{{\bf Equational axioms associated with finite automata 
for fixed point operations
in cartesian categories }}

\author{Zolt\'an \'Esik\thanks{Work partially supported by
the National Foundation of Hungary for Scientific Research (OTKA), 
grant no. ANN 110883.}\\
Dept. of Computer Science\\
University of Szeged\\
Hungary}

\date{February 12, 2015}

\maketitle

\begin{abstract}
The axioms of iteration theories, or iteration categories,
 capture the equational properties 
of fixed point operations in several computationally significant categories. 
Iteration categories may be axiomatized by the Conway identities and 
identities associated with finite automata. We show that the
Conway identities and the identities 
associated with the members of a subclass $\cQ$ of finite automata is complete
for iteration categories iff for every finite simple group $G$ there 
is an automaton $\bQ \in \cQ$ such that $G$ is a quotient of a group 
in the monoid $M(\bQ)$ of the automaton $\bQ$. We also prove 
a stronger result that concerns identities associated 
with finite automata with a distinguished initial state.
\end{abstract}

\section{Introduction}

Most computer scientists are aware of the fact that just about every aspect 
of their study is related to fixed points. The importance of fixed points 
is due to the fact that the semantics of recursion and iteration is usually 
captured by fixed points of functions, functors, or other constructors. 
Fixed point operations have been widely used in automata and formal language
theory and their generalizations, in the semantics of programming languages, 
abstract data types, process algebra and concurrency, rewriting, 
programming logics and verification, complexity theory, and many other fields.
The study of the equational properties of the fixed point operations 
has a long history. Some early references are 
\cite{Bekic,DeBakkerScott,Elgot,Esikaxioms,Niwinski,Plotkin,Wrightetal}. 
A general study of the equational properties of fixed point operations 
has been provided by \emph{iteration theories}, introduced in 1980 
independently in \cite{BEW1} and \cite{Esikaxioms}. Main (but somewhat
outdated) references to iteration theories are \cite{BEbook} and \cite{BEtcs}.
The results obtained in \cite{BEbook,Espark,EsLabella,PlotkinSimpson} show that 
iteration theories give a full account of the equational properties 
of fixed point operations.

An iteration theory is a Lawvere theory \cite{Lawvere} equipped with 
a fixed point (or dagger or iteration) operation, subject
to a set of equational axioms. A Lawvere theory is 
a cartesian category such that each object is 
the $n$-fold coprouct of a generator object with itself for some 
integer $n \geq 0$.  
In this paper, just as in \cite{BEccc,EsIterationCategories,PlotkinSimpson}, 
instead of theories,
we will consider the slightly more general cartesian categories
and refer to iteration theories as iteration categories.

Iteration categories (or rather, theories) were axiomatized in \cite{Esikaxioms} 
by the `Conway identities' (terminology from \cite{BEbook})
and the `commutative identities'. The commutative identities
are essentially the same as the (`vector forms' of the) identities associated 
with finite automata in \cite{Esgroup}. Since each finite 
group may be seen as an automaton, there is also an identity 
associated with each finite group.

In \cite{Esgroup}, it has been shown that the 
Conway identities, together with a collection of the
group identities associated with the groups in a 
subclass $\cG$ of the finite groups is complete 
for iteration categories iff every finite
simple group is a divisor, i.e., a homomorphic 
image of a subgroup, of a group in $\cG$. This result
was derived from the fact that for every finite 
automaton $\bQ$, the identity associated with $\bQ$ 
holds in all Conway theories satisfying the identity
associated with every simple group which divides the 
monoid of $\bQ$. In Theorem~\ref{thm-new}, 
we prove the converse of this fact: if a Conway category satisfies the 
identity associated with a finite automaton $\bQ$,
then it satisfies all identities associated with the 
(simple) group divisors of the monoid of $\bQ$.
The crucial new part of the argument for this result is contained 
in the proof of Lemma~\ref{lem-extension} that shows that if 
the identity associated with an automaton $\bQ$ holds in a Conway category,
then so does each identity associated with any input extension of $\bQ$. 
Using Theorem~\ref{thm-new}, we conclude in Theorem~\ref{thm-main2} that the
Conway identities and the collection of the identities 
associated with a subclass $\cQ$ of finite automata is complete
for iteration categories iff for every finite  
simple group $G$ there is 
an automaton $\bQ \in \cQ$ such that $G$ divides the 
monoid of $\bQ$. This theorem strengthens a result 
proved in \cite{EsIterationCategories}. 
Actually we prove an even stronger result (Theorem~\ref{thm-new2}) 
that concerns identities 
associated with finite automata with a distinguished initial state.

\section{Iteration categories}

In any category $\C$, we will denote the composition of morphisms 
$f: A \to B$ and $g : B \to C$  by $g \circ f: A \to C$. The identity
morphism corresponding to object $A$ will be denoted $\1{A}$. 

We will consider \emph{cartesian categories} $\C$ with explicit products. 
Thus we assume that for any family of objects $C_i$, 
$i \in [n] = \{ 1,\ldots, n \}$, 
there is a fixed product diagram 
\begin{eqnarray*}
\pi_{i}^{C_1 \x \cdots \x C_n}:C_1 \x \cdots \x C_n &\to& C_i, \quad i \in [n]
\end{eqnarray*}
with the usual universal property. When $f_i : A \to C_i$, $i \in [n]$ is a family of morphisms, the unique mediating morphism $f: A \to C_1 \x \cdots \x C_n$ 
with $\pi^{C_1\x \cdots \x C_n}_i \circ f = f_i$ for all $i \in [n]$ will be denoted 
$\le f_1,\ldots,f_n\re$. This morphism is
 called the {\em tupling} of the $f_i$. 
In particular, when $n = 0$, the empty tuple is the unique morphism 
$!_A: A \to T$, where $T$ is the fixed terminal object.    
We will assume that product is `associative on the nose',
so that  $A \x (B \x C) = (A \x B) \x C$, $A \x T = A = T \x A$,
$\langle \1{A},!_b\rangle = \pi^{A\x B}_1$  etc.
for all objects $A,B,C$. 
In particular, we will assume that for each object $A$ the 
projection morphism $\pi_1^A : A \to A$ is the identity 
morphism $\1{A}$. It follows that $\le f \re = f$ 
for all $f: A \to B$.

In the sequel, we will call tuplings of projections {\em base morphism}. 
Note that any base morphism 
$$\langle \pi^{A_1\x \cdots \x A_n}_{i_1},
\ldots, 
\pi^{A_1\x \cdots \x A_n}_{i_m}
\rangle:
A_1\x \cdots \x A_n \to A_{i_1}\x \cdots \x A_{i_m}$$ with $i_1,\ldots,i_m\in [n]$ corresponds to the
function $\rho: [m]\to [n]$ mapping $j \in [m]$ to $i_j \in [n]$. 
For example, for each $A$, the \emph{diagonal morphism}
$\Delta_{A^n} = \langle \1{A},\ldots,\1{A}\rangle : A \to A^n$ corresponds to the unique 
function $[n] \to [1]$.
We call a base morphism 
$A_1 \x \cdots \x A_n \to A_{1\rho} \x \cdots \x A_{n\rho}$ 
corresponding to a permutation
$\rho: [n] \to [n]$ a {\em base permutation}. 

In any cartesian category $\C$, we define 
\begin{eqnarray*}
f \x g &=& \le f \circ \pi_1^{A \x B}, \ g \circ \pi_2^{A \x B} \re : A \x B \to C \x D
\end{eqnarray*}
for all $f: A \to C$, $g: B \to D$. Clearly, $\x$ is
a bifunctor $\C \x \C \to \C$.

Suppose that $\C$ is a cartesian category.
A {\em dagger operation} on $\C$ maps a morphism
$f :A \times C \to A$ to a morphism $f^\dagger : C \to A$.

The {\em Conway identities} are the {\em parameter} (\ref{eq-param}), 
{\em double dagger} (\ref{eq-dd})
and {\em composition identities} (\ref{eq-comp}) given below.   \begin{eqnarray}
\label{eq-param}
(f\circ (\1{A} \x g))\dr &=& f\dr \circ g,
\end{eqnarray}
all $f:A \x B \to A,\ g:C\to B$,
\begin{eqnarray}
\label{eq-dd}
f^{\dagger\dagger} &=& (f \circ (\Delta_{A^2} \x \1{C}))\dr,
\end{eqnarray}
where $f: A \x A \x C \to A$.
\begin{eqnarray}
\label{eq-comp}
(f\circ \le g, \pi_2^{A \x C}\re)\dr 
&=&
f \circ \le (g \circ \le f, \pi_2^{B \x C} \re)\dr,\ \pi_2^{B \x C}\re,
\end{eqnarray}
for all $f:B \x C \to A$, $g:A \x C \to B$. 
Note that the {\em fixed point identity} 
\begin{eqnarray}
\label{eq-fixed point}
f \dr 
&=& 
f\circ \le f\dr,\1{C}\re,\quad f:A\x C \to A
\end{eqnarray}
is a particular subcase of the composition identity,
and the identity 
\begin{eqnarray}
\label{eq-left zero}
(f \circ \pi_2^{A \times C})^\dagger &=& f,\quad f: C\to A
\end{eqnarray}
is a particular subcase of the fixed point identity (\ref{eq-fixed point}), while 
\begin{eqnarray}
\label{eq-right zero}
(f\circ \pi_2^{A \times B \times C})^\dagger &=& f^\dagger \circ \pi^{B\times C}_1,\quad f: A \times B \to A
\end{eqnarray}
is an instance of the parameter identity (\ref{eq-param}).
The parameter, double dagger and composition 
identities are sometimes called 
the identities of {\em naturality}, {\em diagonality} and {\em dinaturality}. 

\begin{deff} {\rm \cite{BEbook}}
A {\em Conway category} is a cartesian category equipped with 
a dagger operation satisfying the Conway identities.
\end{deff}

Conway categories satisfy several other non-trivial identities 
including the {\em Beki\'{c} identity} 
\cite{Bekic} (called the {\em pairing identity} 
in \cite{BEbook})
\begin{eqnarray}
\label{eq-pairing}
\le f, g \re \dr &=& \le f \dr \circ \le h\dr,\1{C}\re,\ h\re \dr,
\end{eqnarray}
or its `dual form'
\begin{eqnarray}
\label{eq-pairing dual}
\le f, g \re \dr &=& \le k^\dagger,\ \overline{g}^\dagger \circ \langle k^\dagger, \1{C}
\rangle \rangle,
\end{eqnarray}
for all $f: A \x B \x C \to A$ and $g: A \x B \x C \to B$, 
where 
\begin{eqnarray*}
h &=& g \circ \le f\dr, \1{B \x C}\re: B\x C \to B\\
k &=& f \circ \le \pi^{A \x C}_1, \overline{g}^\dagger, \pi^{A\x C}_2\re: A \x C \to A
\end{eqnarray*}
with 
\begin{eqnarray*}
\overline{g} &=& g \circ (\langle \pi^{A \x B}_2,\pi^{A\x B}_1\rangle \x \1{C}) : B\x A \x C \to B.
\end{eqnarray*}
By combining the pairing identity (\ref{eq-pairing})
with the identities (\ref{eq-left zero}) and (\ref{eq-right zero}),
we obtain
\begin{eqnarray}
\label{eq-sep}
\langle f\circ (\1{A} \x !_B \x \1{C}), g \circ (!_A \x \1{B \x C})^\dagger 
&=& \langle f^\dagger,g^\dagger\rangle,
\end{eqnarray}
for all $f: A\x C \to A$ and $g: B \x C \to B$. 
Also,
\begin{eqnarray}
\label{eq-c1}
\pi^{A \x B}_1 \circ \langle f \circ (\1{A} \x !_B \x \1{C}), g\rangle^\dagger 
&=& f^\dagger,
\end{eqnarray}
for all $f: A \x C \to A$ and $g: A \x B \x C \to B$, and
\begin{eqnarray}
\label{eq-c2}
\langle f, g \x !_{B \x C}  \rangle^\dagger &=& 
\langle f\circ (\langle \1{A}, g \rangle \x !_B \x \1{C}), g \x !_{B \x C} \rangle^\dagger\\
\label{eq-c22}
&=&\langle (f \circ (\langle \1{A}, g\rangle\x \1{C}))^\dagger, g \circ (f \circ (\langle \1{A}, g\rangle\x \1{C}))^\dagger\rangle
\end{eqnarray} 
for all $f: A \x B \x C \to A$ and $g: A\to B$.

We will also make use of the {\em permutation identity}
that holds in all Conway categories:
\begin{eqnarray}
\label{eq-perm}
(\pi \circ f \circ (\pi^{-1} \x \1{C}))\dr &=& \pi \circ f\dr,
\end{eqnarray}
for all $f:A_1\x \cdots \x A_n\x C \to A_1 \x \cdots \x A_n$ and all base permutations 
$$\pi:A_1\x \cdots \x A_n \to A_{1\pi} \x \cdots \x A_{n\pi}.$$ 
Note also that, by repeated applications of (\ref{eq-pairing}), it follows that 
\begin{eqnarray} 
\label{eq-ndagger}
f^{\dagger\cdots \dagger} &=& (f \circ (\Delta_{A^n}\x \1{C}))^\dagger
\end{eqnarray} 
for all $f: A^n \x C \to A$, where dagger appears $n\geq 1$ times on the left-hand side.

A concrete description of the free Conway categories 
and a characterization of the valid identities 
of Conway categories together with a PSPACE decision 
algorithm were given in \cite{BerEs}.
In the sequel, we will also make use of the following fact.

\begin{lem}
\label{lem-adding id}
Suppose that $f_i: A^{1+n} \times C \to A$ for each $i \in [n]$ in a Conway category $\C$.
Let 
\begin{eqnarray*}
g &=& 
\langle f_1 \circ ( \langle \pi^{A^n}_1, \1{A^n} \rangle \os \1{C}),
\ldots,
f_n \circ (\langle \pi^{A^n}_n ,\1{A^n} \rangle \os \1{C}) \rangle: A^n \times C \to A^n.
\end{eqnarray*}
Then 
\begin{eqnarray*}
g^\dagger &=& \langle f_1^\dagger,\ldots,f_n^\dagger \rangle^\dagger.
\end{eqnarray*}
\end{lem}  

{\sl Proof.} Define $h: A^{2n}\times C \to A^n$ by 
\begin{eqnarray*}
h &=& \langle f_1 \circ (\pi^{A^n}_1 \os \1{A^n\times C}),
\ldots,
f_n \circ (\pi^{A^n}_n \os \1{A^n \times C})\rangle.
\end{eqnarray*}
Then, by the double dagger identity (\ref{eq-dd}) and repeated applications of (\ref{eq-sep}),
\begin{eqnarray*}
g^\dagger &=& (h\circ (\langle \1{A^n},\1{A^n}\rangle \times \1{C}))^\dagger \\
&=& h^{\dagger\dagger}\\
&=& \langle f_1^\dagger,\ldots,f_n^\dagger\rangle^\dagger. \eop
\end{eqnarray*}

The Conway identities are quite powerful, for example, a general 
`Kleene theorem' holds in all Conway categories, and both the 
soundness and relative completeness of Hoare's logic can be 
proved just from the Conway identities, cf. \cite{BEbook}. 
However, they are not complete for the equational theory 
of the fixed point operations commonly used in computer science.
The missing ingredient is formulated by the notion of 
identities associated with finite automata, called automaton identities,
or identities associated 
with finite groups, called group identities. 
In order to define these identities, 
we need to introduce some definitions and notation.

Suppose that $\bQ = (Q,Z,\cdot)$ is a \emph{(deterministic) finite automaton}, where $Q$ 
is the finite nonempty set of states, $Z$ is the finite nonempty set 
of input letters, and $\cdot: Q \times Z \to Q$ is the (right) action 
of $Z$ on $Q$. The action is extended in the usual way to a right action 
of $Z^*$ on $Q$, where $Z^*$ denotes the free monoid of all 
words over $Z$ including the empty word $\epsilon$. When $q\in Q$ 
and $u \in Z^*$, we will usually write just $qu$ for $q \cdot u$.

When $u \in Z^*$, we call the function $Q \to Q$ defined by 
$q \mapsto qu$ for all $q \in Q$ the \emph{function induced by $u$}
and denote it by $u^\bQ$. These functions form a (finite) monoid 
$M(\bQ)$ whose multiplication is given by $u^\bQ\cdot v^\bQ = (uv)^\bQ = 
v^\bQ \circ u^\bQ$, for all $u,v \in Z^*$. 

Suppose that $\bQ = (Q,Z,\cdot)$ and $\bQ' = (Q, Z',\cdot)$ are finite 
automata with the same set of states. We say that $\bQ'$ 
is an \emph{extension of $\bQ$} if $Z \subseteq Z'$, 
$a^{\bQ} = a^{\bQ'}$ for all $a \in Z$, 
and if for each letter $b \in Z' \setminus Z$ there is some word 
$u \in Z^*$ with $b^{\bQ'} = u^{\bQ}$. Moreover, we say that
$\bQ'$ is a \emph{restriction of $\bQ$} if $Z' \subseteq Z$
and $b^{\bQ'} = b^{\bQ}$ for all $b \in Z'$. Note that
when $\bQ$ is a restriction of $\bQ'$ this does not imply 
that $\bQ'$ is an extension of $\bQ$. Also, when 
$\bQ'$ is an extension of $\bQ$ then $M(\bQ') = M(\bQ)$, and 
if $\bQ'$ is a restriction of $\bQ$ then $M(\bQ')$ is a submonoid 
of $\bQ$.  Each finite automaton $\bQ = (Q,Z,\cdot)$ has an 
extension $\bQ' = (Q,Z',\cdot)$ such that for each $u \in Z^*$ there is 
some $b\in Z'$ with $b^{\bQ'} = u^\bQ$, 
and for any two finite automata $\bQ,\bQ'$ with the same 
state set $Q$, if $M(\bQ) = M(\bQ')$
then, up to a renaming of the input letters, 
$\bQ'$ is a restriction of an extension of $\bQ$.

Suppose that $\bQ = (Q,Z,\cdot)$ is a finite automaton with
$Q = \{q_1,\ldots,q_n\}$ and $Z = \{a_1,\ldots,a_m\}$, say.
Let $\C$ be a Conway category and $f: A^m \times C \to A$
in $\C$. For each $i \in [n]$, let
$$\rho_i^{\bQ,A} = 
\langle \pi_{ia_1}^{A^n},\ldots,\pi_{ia_m}^{A^n}\rangle : 
A^n \to A^m,$$
\emph{where we identify each state $q_i$ with the integer $i$.}
Moreover, let 
\begin{eqnarray*}
f^{\bQ,A} &=&f \parallel (\rho_1^{\bQ,A},\ldots,\rho_n^{\bQ,A}): 
A^n \times C \to A^n
\end{eqnarray*}
be the morphism such that for all $i \in [n]$,
\begin{eqnarray*}
\pi^{A^n}_i \circ f^{\bQ,A}
&=& 
f\circ (\rho_i^{\bQ,A} \times \1{C}): A^n \times C \to A.
\end{eqnarray*}

\begin{deff}
We say that the identity $\Gamma(\bQ)$ associated with $\bQ$ holds 
in a Conway category $\C$ if for all $f: A^m \times C \to A$,
\begin{eqnarray*}
(f^{\bQ,A})^\dagger &=& 
\Delta_{A^n} \circ (f \circ(\Delta_{A^m}\times \1{C}))^\dagger.
\end{eqnarray*}
Suppose that a state $q = q_i$ of $\bQ$ is distinguished,
so that $(\bQ,q)$ is an \emph{initialized finite automaton}.
 We say that 
the identity $\Gamma(\bQ,q)$ holds in $\C$ if 
\begin{eqnarray*}
\pi^{A^n}_i \circ (f^{\bQ,A})^\dagger 
&=& 
(f \circ(\Delta_{A^m}\times \1{C}))^\dagger
\end{eqnarray*}
for all $f: A^m \times C \to A$.
\end{deff}
By the permutation identity which holds in all Conway categories,
this definition does not depend on the enumeration of the states
of $\bQ$. It also does not depend on the enumeration of the input
letters, since we may replace $f$ with a morphism $g = f \circ \pi$,
where $\pi: A^m \to A^m$ is a base permutation.
It is clear that $\Gamma(\bQ)$ holds in a Conway category $\C$ 
iff $\Gamma(\bQ,q)$ holds in $\C$ for all states $q$ of $\bQ$.

\begin{deff} \cite{Esikaxioms}
An \emph{iteration category} is a Conway category satisfying all
identities associated with finite automata, or initialized finite 
automata.
\end{deff}

Only iteration categories generated by a single object were 
considered in \cite{Esikaxioms}, called generalized iterative theories,
but the generalization is straightforward. See also \cite{BEccc,EsIterationCategories}.

\begin{remark}
The first axiomatization of iteration categories (or rather, theories) 
involved some of the Conway identities and the commutative identities,
cf. \cite{BEbook,Esikaxioms}. 
In cartesian categories, the identity associated with the a finite 
automaton is clearly an instance of the commutative identities.
Also, in Conway categories, each instance of the commutative identities 
is easily seen to be equivalent to the identity associated with an automaton.
\end{remark}

An example of an iteration category is the category $\Cpo$
of complete partial orders and continuous (or monotone) functions,
equipped with the (parameterized) least fixed point operation
as dagger. 
It is known that an identity involving the cartesian category operations and 
dagger holds in $\Cpo$ iff it holds in iteration categories. 
For this fact and other completeness results see e.g. \cite{BEbook,BEtcs,Esikaxioms,EsLabella,Espark,PlotkinSimpson}.

It is instructive to explain the meaning of some of the above identities 
of iteration categories using functional notation, over the category
$\Cpo$, say, or an appropriate category of sets with structure and structure 
preserving maps, equipped with dagger. The meaning of the fixed point identity is, of course, that for all $f: A \x C \to A$ and $y \in C$,
$f^\dagger(y)$ is a (canonical) solution of the fixed point equation 
$x = f(x,y)$. By the parameter identity (\ref{eq-param}), 
if $g: D \to C$, then for all $z\in D$, the canonical
solution of $x = f(x,g(z))$ is $f^\dagger(g(z))$.

The double dagger identity (\ref{eq-dd}) 
asserts that for $f:A \times A \times C \to C$, the canonical solution 
of the equation $x = f(x,x,y)$ may be obtained by first solving $x = f(x,y,z)$ 
to obtain $f^\dagger(y,z)$, and then solving $x = f^\dagger(x,z)$.

Suppose that $\bQ$ is the two-state automaton with $4$ input letters 
inducing the $4$ possible state maps. Then the identity $\Gamma(\bQ)$ associated 
with $\bQ$ asserts that for all $f: A^4 \x C \to A$, 
both components of the canonical solution of the system
\begin{eqnarray*}
x &=& f(x,y,x,y,z)\\
y &=& f(y,x,x,y,z)
\end{eqnarray*}
are equal to the canonical solution of the single equation 
$x = f(x,x,x,x,z)$.
(Here, the input letters are enumerated so that the first letter
induces the identity function, the second induces the swapping 
of the two states, while the last two letters induce the two constant maps.)

As another example, let $\bQ$ be the automaton with $n$ states and a single 
letter inducing a cyclic permutation. Then $\Gamma(\bQ)$ asserts that 
each component of the canonical solution of the system 
\begin{eqnarray*}
x_1 &=& f(x_2,y)\\
&\vdots &\\
x_{n-1} &=& f(x_n,y)\\
x_n &=& f(x_1, y)
\end{eqnarray*}
for $f: A \x C \to A$
is equal to the solution of the single equation $x = f(x,y)$.
In fact, in Conway categories, $\Gamma(\bQ)$ is equivalent to 
the \emph{power identity} $(f^n)^\dagger = f^\dagger$, where $f: A \x C \to A$,
and the powers of $f$ are defined by $f^1 = f$ and 
$f^n = f \circ \langle f^{n-1}, \pi_2^{A \times C}\rangle$, for all $n > 1$. 

\section{Main results}

We may view each finite monoid $M$ as the finite automaton
$(M,M,\cdot)$, where the action is given by monoid multiplication.
The identity corresponding to this automaton is called 
the \emph{(monoid) identity associated with the monoid $M$}.
It is denoted $\Gamma(M)$. 
When $M$ is a group, $\Gamma(M)$ is called a \emph{group identity}.

When $M$ is a finite monoid, by a \emph{group in $M$} we shall mean
a subsemigroup of $M$ which is a group. Thus, a group in $M$ 
is not required to contain the identity element of $M$, its unit element 
may be any idempotent element of $M$. However, if $M$ is 
itself a group, then a group in $M$ is just a subgroup of $M$.
Moreover, we say that a (finite) group $G$ \emph{divides} $M$ if $G$ 
is a quotient of a group in $M$. 

We recall that a finite group $G$ is \emph{simple} if it is nontrivial 
and its only nontrivial normal subgroup is $G$ itself.

The following result was proved in \cite{Esgroup} (see Theorem 16.1).

\begin{thm}
\label{thm-old}
Suppose that $\C$ is a Conway category and $\bQ$ is a finite automaton.
If $\C$ satisfies the identity associated with each simple group
divisor of $M(\bQ)$, then it satisfies the identity associated 
with $\bQ$.
\end{thm}

Our aim is to prove the following converse of Theorem~\ref{thm-old}.

\begin{thm}
\label{thm-new}
If a Conway category satisfies the identity $\Gamma(\bQ)$ associated with a 
finite automaton $\bQ$, then it satisfies the group identity associated
with any simple group divisor of $M(\bQ)$. 
\end{thm}

In our argument, we will make use Lemma~\ref{lem-extension}, 
proved below, and some results from \cite{Esgroup}: 

\begin{thm} {\rm \cite{Esgroup}}
\label{thm-monoid}
Suppose that $M$ is a finite monoid and $\C$ is a Conway category.
Then the following are equivalent.
\begin{itemize}
\item $\C$ satisfies the identity $\Gamma(M)$ associated with $M$.
\item $\C$ satisfies the identity associated with 
every group in $M$.
\item $\C$ satisfies the identity associated with every 
group dividing $M$.
\item $\C$ satisfies the identity associated with every 
simple group dividing $M$.
\end{itemize}
\end{thm}

Indeed, the fact that the first condition implies the second is 
covered by Corollary 11.3 of \cite{Esgroup}, and the fact that 
the second and fourth conditions are equivalent is proved 
in Proposition 13.5 of \cite{Esgroup}. It then follows
that the last three conditions are all equivalent.
Finally, the last condition implies the first 
condition by Theorem 16.1 of \cite{Esgroup}. 

Actually we will prove a stronger version of Theorem~\ref{thm-new}.
We say that an initialized finite automaton $(\bQ,q)$ is \emph{initially 
connected} if every state of $\bQ$ is accessible from $q$ by some 
word, i.e., when each state of $\bQ$ may be written as $qu$ 
for some word $u$.

\begin{thm}
\label{thm-new2}
If a Conway category satisfies the identity $\Gamma(\bQ,q)$ 
associated with an initialized finite automaton $(\bQ,q)$, and 
if $(\bQ,q)$ is initially connected, then it satisfies the group identity 
associated with every simple group divisor of $M(\bQ)$. 
\end{thm}

\section{Proof of the main results}

In this section, we provide a proof of Theorem~\ref{thm-new}
and Theorem~\ref{thm-new2}.

\begin{lem}
\label{lem-extension}
If an identity associated with a finite automaton holds in a Conway category,
then so does the identity associated with any extension of the automaton. 
\end{lem}

{\sl Proof.}
Let $\C$ be a Conway category. Suppose that the identity  $\Gamma(\bQ)$ 
associated with the automaton $\bQ$ holds in $\C$, where $\bQ = (Q,Z,\cdot)$.
Let $Q = \{q_1,\ldots,q_n\}$ and $Z = \{a_1,\ldots,a_m\}$, say.
Below we will identify state $q_i$ with the integer $i$.

First we consider the extension of $\bQ$ with a letter $c$ inducing
the identity function, also treated in \cite{EsIterationCategories}.
 So let $\bQ' = (Q,Z\cup\{c\},\cdot)$  
be the extension of $\bQ$, where $c$ induces the identity function 
$Q \to Q$. Let us enumerate the elements of $Z \cup\{c\}$ as 
$c,a_1,\ldots,a_m$ (we may use any order), 
and suppose that $f:A^{1+m}\x C \to A$
in $\C$. First we explain the argument using functional
notation, where $\C$ is appropriate. The system of fixed point
equations corresponding to $\bQ'$ with respect to $f$ is 
\begin{eqnarray*}
x_1 &=& f(x_1,x_{1a_1},\ldots,x_{1a_m},y)\\
&\vdots &\\
x_n &=& f(x_n,x_{na_1},\ldots,x_{na_m},y),
\end{eqnarray*}
where for all appropriate $i,j$, $ia_j$ is the state entered by 
$\bQ$ or $\bQ'$ from state $i$ upon receiving the letter $a_j$.
By Lemma~\ref{lem-adding id}, this system can be solved in two steps.
First, let 
\begin{eqnarray*}
x_1 &=& f^\dagger(x_{1a_1},\ldots,x_{1a_m},y)\\
&\vdots&\\
x_n &=& f^\dagger(x_{na_1},\ldots,x_{na_m},y)
\end{eqnarray*}
and then solve this system, which is just the 
system of fixed point equations corresponding to $\bQ$ 
with respect to $f^\dagger: A^n \x C \to A$.
Since the identity associated with $\bQ$ holds, 
each component of the solution of this latter system 
is $h^\dagger(y)$, where $h(x,y) = f^\dagger(x,\ldots,x,y)$.
But by (\ref{eq-ndagger}), $h^\dagger(y) = k^\dagger(y)$,
where $k(x,y) = f(x,\ldots,x,y)$.

More formally, we have
\begin{eqnarray*}
\rho^{\bQ',A}_i &=& \langle \pi^{A^n}_i,\rho^{\bQ,A}_i\rangle : A^n \to A^{1+m}
\end{eqnarray*}
for all $i \in [n]$, so that 
\begin{eqnarray*}
f^{\bQ',A} &=&
f\parallel (\rho^{\bQ',A}_1,\ldots,\rho^{\bQ',A}_n)\\
&=&
\langle 
f \circ (\langle \pi^{A^n}_1,\rho^{\bQ,A}_1\rangle \times \1{C}),
\ldots, 
f \circ (\langle \pi^{A^n}_n,\rho^{\bQ,A}_n\rangle \times \1{C}) \rangle\\
&=& 
\langle 
f \circ (\1{A} \times \rho^{\bQ,A}_1 \times \1{C}) \circ 
(\langle \pi^{A^n}_1,\1{A^n} \rangle \times \1{C}),
\ldots, \\
&&f \circ (\1{A} \times \rho^{\bQ,A}_n \times \1{C}) \circ
(\langle \pi^{A^n}_n,\1{A^n} \rangle \times \1{C})
\rangle: A^n \x C \to A^n.
\end{eqnarray*} 
Thus,
\begin{eqnarray*}
(f^{\bQ',A})^\dagger 
&=& 
\langle f^\dagger \circ( \rho^{\bQ,A}_1 \times \1{C}),\ldots,
f^\dagger \circ( \rho^{\bQ,A}_n \times \1{C})\rangle^\dagger,
\end{eqnarray*}
by Lemma~\ref{lem-adding id} and the parameter identity (\ref{eq-param}),
\begin{eqnarray*}
&=& 
((f^\dagger)^{\bQ,A})^\dagger\\
&=&
\Delta_{A^n} \circ (f^\dagger\circ (\Delta_{A^m} \times \1{C}))^\dagger,
\end{eqnarray*}
since $\Gamma(\bQ)$ holds in $\C$, 
\begin{eqnarray*}
&=& 
\Delta_{A^n}\circ (f\circ (\Delta_{A^{1+m}}\times \1{C}))^\dagger,
\end{eqnarray*}
by (\ref{eq-ndagger}). 
This proves that the identity $\Gamma(\bQ')$ associated with $\bQ'$ holds in $\C$.

Next let $\bQ' = (Q,Z \cup\{c\},\cdot)$ be the extension of $\bQ$ with an 
input letter $c$ inducing the same function as the two-letter word $ab$, 
for some $a,b \in Z$. We prove that $\Gamma(\bQ')$ holds in $\C$. 
Let $a = a_m$ and $b = a_{j_0}$, say. 
Moreover, let 
$g: A^{m+1}\times C \to A$, and define:
\begin{eqnarray*}
f_1 &=&
g \circ (\pi_1^{A^2}\os\cdots \os \pi_1^{A^2}\os \1{A^2 }\os \1{C}): A^{2m}\os C \to A\\
f_2 &=& 
\pi^{A^{2m}}_{2(j_0-1)+1} \os !_{C}:  A^{2m}\os C \to A,
\end{eqnarray*}
where $\pi_1^{A^2}$ appears $m-1$ times in the definition of $f_1$.
Let $f = \langle f_1,f_2\rangle: B^m \os C \to B$, where $B = A^2$. Since $\Gamma(\bQ)$
holds in $\C$, we have that each component of 
$$ (f^{\bQ,B})^\dagger: C \to B^n$$
is equal to 
$$(f\circ (\Delta_{B^m} \os \1{C}))^\dagger: C \to B.$$

Let $\alpha = \pi_1^{A^2}\times \cdots \times \pi_1^{A^2}: B^n \to A^n$.
Thus, 
\begin{eqnarray}
\label{eq-1}
\alpha \circ (f^{\bQ,B})^\dagger
&=& 
\Delta_{A^n}\circ \pi_1^{A^2} \circ (f \circ (\Delta_{B^m} \os\1{C}))^\dagger : C \to A^n.
\end{eqnarray}
However, we will prove that the left-hand side of (\ref{eq-1}) is 
$$(g^{\bQ',A})^\dagger$$
and the right-hand side of (\ref{eq-1}) is 
$$
\Delta_{A^n} \circ (g \circ (\Delta_{A^{m+1}} \os \1{C}))^\dagger,
$$
so that 
\begin{eqnarray*}
(g^{\bQ',A})^\dagger
&=&
\Delta_{A^n} \circ (g \circ (\Delta_{A^{m+1}} \os \1{C}))^\dagger.
\end{eqnarray*}

To see informally that this holds, suppose that $\C$ 
is an appropriate category of sets with structure and structure preserving maps.
Consider the system of fixed point equations associated with $\bQ$ 
with respect to $f$, 
\begin{eqnarray*}
x_i &=& f_1(x_{ia_1},y_{ia_1}, \ldots, x_{ia_m},y_{ia_m},z) = 
        g(x_{ia_1},\ldots,x_{ia_m},y_{ia_m},z)\\
y_i &=& f_2(x_{ia_1},y_{ia_1}, \ldots, x_{ia_m},y_{ia_m},z) = 
        x_{ia_{j_0}},
\end{eqnarray*}
where $i \in [n]$. Here, $x_i$ and $y_i$ range over $A$
and $z$ ranges over $C$.
The left-hand side of (\ref{eq-1}) corresponds to 
the canonical solution of this system for the variables $x_1,\ldots,x_n$.
Now to solve this system, by the Conway identities, we may 
substitute $x_{ia_{j_0}}$ for $y_i$ on the right-hand side 
of each equation, for all $i \in [n]$,
so that we obtain a new 
system of fixed point equations
\begin{eqnarray*}
x_i &=&  
g(x_{ia_1},\ldots,x_{ia_{m-1}},x_{ia_m},x_{ia_ma_{j_0}},z)\\
y_i &=& x_{ia_{j_0}}.
\end{eqnarray*}
For each $i$, the right-hand side of the equation for $x_i$ 
does not contain any of the variables $y_j$. Thus, we may 
consider the equations for the $x_i$ separately: 
\begin{eqnarray*}
x_i &=& g(x_{ia_1},\ldots,x_{ia_m},x_{iab},z)
\end{eqnarray*}
for all $i \in [n]$. This is exactly the system of equations 
corresponding to the automaton $\bQ'$ with respect to $g$. Thus,
the left-hand side of (\ref{eq-1}) is the canonical solution 
of the system of equations associated with $\bQ'$ with respect to $g$.
 
Now $(f \circ (\Delta_{B^m}\x  \1{C}))^\dagger$ is the canonical solution 
of the system
\begin{eqnarray*}
x &=& f_1(x,y,\ldots,x,y,z) = g(x,\ldots,x,y,z)\\
y &=& f_2(x,y,\ldots,x,y,z) = x,
\end{eqnarray*} 
and the first component of the solution is the canonical solution of 
the single fixed point equation
\begin{eqnarray*}
x &=& g(x,\ldots,x,x,z).
\end{eqnarray*}
Thus, each component of the right-hand side of (\ref{eq-1}) is 
this canonical solution.

Formally, for each $i \in [n]$, we have  
\begin{eqnarray*}
\pi^{B^n}_i \circ f^{\bQ,B} &=& 
\pi^{B^n}_i \circ (f \parallel  (\rho^{\bQ,B}_1,\ldots,\rho^{\bQ,B}_n))\\
&=& 
\langle f_1 \circ 
(\rho^{\bQ,B}_i \os \1{C}), f_2 \circ (\rho^{\bQ,B}_i \os \1{C}) \rangle\\
&=& 
\langle g \circ (\langle \pi^{A^2}_1\circ \pi^{B^{n}}_{ia_1},\ldots,
\pi^{A^2}_1\circ \pi^{B^{n}}_{ia_m},\pi^{A^2}_2\circ \pi^{B^{n}}_{ia_m} \rangle \os \1{C}),\ 
(\pi^{A^2}_1\circ \pi^{B^{n}}_{ia_{j_0}}) \os !_{C} \rangle.
\end{eqnarray*}
It follows using (\ref{eq-c2}) and the permutation identity (\ref{eq-perm}) that
\begin{eqnarray*}
(f^{\bQ,B})^\dagger
&=& h^\dagger,
\end{eqnarray*}
where $h: B^{n}\times C \to B^{n}$ 
with 
\begin{eqnarray*}
\lefteqn{
\pi^{B^n}_i\circ h = }\\
&=& 
\langle 
g \circ (\langle \pi^{A^2}_1\circ \pi^{B^{n}}_{ia_1},\ldots,
\pi^{A^2}_1\circ \pi^{B^{n}}_{ia_m},\pi^{A^2}_2\circ \pi^{B^{n}}_{ia_ma_{j_0}} 
\rangle \os \1{C}),\ 
(\pi^{A^2}_1\circ \pi^{B^{n}}_{ia_{j_0}} ) \os !_{C} \rangle \\
&=& 
\langle
g \circ (\langle \pi^{A^2}_1\circ \pi^{B^{n}}_{ia_1},\ldots,
\pi^{A^2}_1\circ \pi^{B^{n}}_{ia_m},\pi^{A^2}_2\circ \pi^{B^{n}}_{iab}\rangle \os \1{C}),\ 
(\pi^{A^2}_1\circ \pi^{B^{n}}_{ib} ) \os !_{C} \rangle
\end{eqnarray*}
for all $i \in [n]$.
Let $\gamma: A^{2n}\to A^{2n}$ be the base permutation 
corresponding to the function $[2n]\to [2n]$ which maps 
every integer of the form $2(i-1)+ 1$ for $i \in [n]$ to $i$, and every integer 
of the form $2i$ for $i \in [n]$ to $n + i$. Then for each $i \in [n]$,  
the $i$th component of $\overline{h} = \gamma^{-1} \circ h \circ (\gamma \times \1{C})
: A^{2n} \os C \to A^{2n}$ is the morphism 
\begin{eqnarray*}
g \circ (\langle  \pi^{A^{n}}_{ia_1},\ldots,
 \pi^{A^{n}}_{ia_m},\pi^{A^{n}}_{iab} \rangle 
\os !_{A^{n}}\os \1{C}) 
&=& 
g \circ (\rho^{\bQ',A}_i \times !_{A^n}\os  \1{C}): A^{2n}\x C \to A,
\end{eqnarray*} 
so that the first $n$ components of $\overline{h}$ determine the morphism 
\begin{eqnarray*}
(g \parallel (\rho^{\bQ',A}_1,\ldots,\rho^{\bQ',A}_m,\rho^{\bQ',A}_{m+1}))
\circ (\1{A^n}\os !_{A^n} \os\1{C}) = g^{\bQ',A} \circ (\1{A^n} \x !_{A^n} \x \1{C}): A^{2n}\x C \to A^n.
\end{eqnarray*}
Thus, by (\ref{eq-c1}) and the permutation identity, the tupling of the 
first $n$ components of $\overline{h}^\dagger$ and hence the morphism on the 
left-hand side of (\ref{eq-1}) is exactly 
$(g^{\bQ',A})^\dagger: C \to A^n$.

Now $f \circ (\Delta_{B^m} \os \1{C}) : B \os C \to B$ is
$$\langle g \circ (\Delta_{A^m} \os \1{A}\os \1{C}), \pi_1^{A^2}\os !_C \rangle,$$
and thus, it follows  using (\ref{eq-c22}) that 
 \begin{eqnarray*}
\pi_1^{A^2} \circ (f \circ (\Delta_{B^m} \os\1{C}))^\dagger
&=& 
(g \circ (\Delta_{A^{m+1}} \os \1{C}))^\dagger.
\end{eqnarray*}
This proves that the morphism on the right-hand side of (\ref{eq-1}) is 
$\Delta_{A^n}\circ (g \circ (\Delta_{A^{m+1}} \os \1{C}))^\dagger$.

It follows now that if $\bQ'$ is an arbitrary extension of $\bQ$, then 
the identity $\Gamma(\bQ')$ holds in $\C$.
\eop

\begin{prop}
Suppose that $\C$ is a Conway category and $\bQ$ is a finite automaton.
If the identity associated with $\bQ$ holds in $\C$, then so does the 
identity associated with $M(\bQ)$.
\end{prop}

{\sl Proof.} Suppose that $\bQ = (Q,Z,\cdot)$, and let $M = M(\bQ)$.
Consider the transformation monoid \cite{Eilenberg} $(Q,M)$ viewed as an 
automaton $\bQ' = (Q,M,\cdot)$ in the usual way,
so that $q \cdot u^{\bQ'}  = qu$, for all $q \in Q$ and $u\in Z^*$.
 Similarly,
let $\bQ_M = (M,M,\cdot)$ be the automaton corresponding 
to the monoid $M$.
It is well-known that $\bQ_M$ is isomorphic to a subautomaton of a direct power of $\bQ'$.
If the identity associated with $\bQ$ holds in $\C$, then so does the 
identity associated with $\bQ'$, by Lemma~\ref{lem-extension}. Since 
the automaton $\bQ_M$ associated with $M$ is isomorphic to a subautomaton 
of a direct power of $\bQ_M$, it follows using 
Lemma 11.2 and Corollary 12.2 of \cite{Esgroup}
that the identity $\Gamma(\bQ_M)$ associated with 
$\bQ_M$ also holds.  But the identity associated with $\bQ_M$ is just the 
identity $\Gamma(M)$ associated with $M$.
\eop 

We now complete the proof of Theorem~\ref{thm-new}. 
Suppose that $\C$ is a Conway category and $\bQ$ is a finite automaton
such that the identity associated with $\bQ$ holds in $\C$. 
Then, by the previous proposition, 
the identity associated with $M(\bQ)$ holds in $\C$.
But, by Theorem~\ref{thm-monoid}, this implies that the identity
associated with every (simple) group dividing $M(\bQ)$ holds in $\C$.

We now turn to the proof of Theorem~\ref{thm-new2}. We need the following fact.

\begin{lem}
\label{lem-initial}
Suppose that $\bQ = (Q,Z,\cdot)$ is a finite automaton, $q \in Q$ and $a \in Z$.
If the identity $\Gamma(\bQ,q)$ holds in a Conway category, then so does the 
identity $\Gamma(\bQ,qa)$.
\end{lem}

{\sl Proof.} Let $Q =\{q_1,\ldots,q_n\}$ and $Z = \{a_1,\ldots,a_m\}$.  
Without loss of generality, we may assume that $q = q_1$ and 
$a = a_1$. Let us identify each state $q_i$ with the integer 
$i$. Suppose that $\C$ is a Conway category satisfying 
$\Gamma(\bQ,q)$. In order to prove that $\C$ also satisfies 
$\Gamma(\bQ,qa)$, let $g:A^m \x C \to A$ in $\C$,  
$B = A^2$, and define $f = \langle f_1,f_2\rangle: B^m \x C \to B$
by 
\begin{eqnarray*}
f_1 &=& g \circ(\pi^{A^2}_1 \x \cdots \x \pi^{A^2}_1 \x \1{C})\\
f_2 &=& \pi_1^{A^{2m}} \x !_C. 
\end{eqnarray*}
Since the identity associated with $(\bQ,q)$ holds, we have 
\begin{eqnarray}
\label{eq-xx}
\pi^{B^n}_1 \circ (f^{\bQ,B})^\dagger
&=& (f \circ (\Delta_{B^m} \x \1{C}))^\dagger.
\end{eqnarray}
Let $\gamma$ denote the base permutation $A^{2n}\to A^{2n}$ 
corresponding to the function $[2n]\to [2n]$ given by 
$(2j - 1)\mapsto j$ and $2j \mapsto n+j$, for all $j \in [n]$. 
Let $h = \gamma^{-1} \circ (f^{\bQ,B} \circ (\gamma \x \1{C}))$.
It is a routine calculation to verify that 
\begin{eqnarray*}
h &=& \langle g^{\bQ,A} \circ (\1{A^n} \x !_{A^n} \x \1{C}),
  \langle \pi^{A^n}_{1a_1}, \ldots, \pi^{A^n}_{na_1} \rangle \x  !_{A^n \x C} \rangle: A^{2n} \x C \to A^{2n}.
\end{eqnarray*} 
Now
\begin{eqnarray*}
h^\dagger &=& \langle (g^{\bQ,A})^\dagger, \langle \pi^{A^n}_{1a_1},\ldots,
    \pi^{A^n}_{na_1}\rangle \circ (g^{\bQ,A})^\dagger\rangle: C \to A^{2n}, 
\end{eqnarray*} 
by (\ref{eq-c22}), and thus by the permutation identity,
\begin{eqnarray*}
\pi^{B^n}_1 \circ (f^{\bQ,A})^\dagger
&=& \langle \pi^{A^n}_1 \circ (g^{\bQ,A})^\dagger, \pi^{A^n}_{1a_1} \circ (g^{\bQ,A})^\dagger \rangle: C \to A^2.
\end{eqnarray*}
On the other hand, by a similar argument, 
\begin{eqnarray*}
(f \circ (\Delta_{B^m} \x \1{C}))^\dagger  &=&  
\langle g \circ (\Delta_{A^m} \x !_{A^m} \x \1{C}), \1{A} \x !_{A} \x \1{C}\rangle^\dagger\\
&=& \Delta_{A^2} \circ (g \circ (\Delta_{A^m}\x \1{C}))^\dagger: C \to A^2.
\end{eqnarray*}
Since (\ref{eq-xx}) holds, it follows that  $\pi^{A^n}_1 \circ (f^{\bQ,A})^\dagger = (g \circ (\Delta_{A^m}\x \1{C}))^\dagger$.
\eop

By repeated applications of Lemma~\ref{lem-initial} we obtain:

\begin{cor}
\label{cor-initial}
Suppose that $\bQ$ is an automaton and $q$ is a state of $\bQ$ and $u$ 
is an input word of $\bQ$. If the identity
$\Gamma(\bQ,q)$ holds in a Conway category $\C$, 
then so does the identity $\Gamma(\bQ, qu)$.
Thus, if $(\bQ,q)$ is initially connected,  
then $\Gamma(\bQ)$ holds in a Conway category $\C$
iff $\Gamma(\bQ,q)$ does. 
\end{cor}

\begin{remark} Our proof of Corollary~\ref{cor-initial} 
based on Lemma~\ref{lem-initial} uses essentially
the same technique as the proof of Lemma 4.3 in \cite{PlotkinSimpson}.
\end{remark}

Theorem~\ref{thm-new2} is now an immediate corollary of Theorem~\ref{thm-new}
and Corollary~\ref{cor-initial}.

\section{Some corollaries and open problems}

Combining Theorem~\ref{thm-old} and Theorem~\ref{thm-new}, 
we immediately have:

\begin{thm}
\label{thm-main2}
Suppose that $\cQ$ is a collection of finite automata 
and let $\cG$ denote the collection of simple groups
dividing the monoid of some automaton in $\cQ$. 
Then a Conway category satisfies the identities 
associated with the automata in $\cQ$ iff it satisfies 
the identities associated with the groups in $\cG$. 
\end{thm}

{\sl Proof.} Suppose that a Conway category $\C$ satisfies 
the identities $\Gamma(G)$ associated with the groups $G$ in $\cG$.
Then by Theorem~\ref{thm-old}, $\C$ satisfies all identities 
$\Gamma(\bQ)$, where $\bQ$ is a finite automaton
such that every simple group divisor of $M(\bQ)$ is
in $\cG$ and thus all the identities associated with the 
finite automata in $\cQ$. 

Suppose now that $\C$ is a Conway category satisfying the identities 
associated with the finite automata in $\cQ$. 
Then by Theorem~\ref{thm-new}, every identity associated with 
the groups in $\cG$ holds in $\C$. \eop 

Similarly, we have:

\begin{thm}
\label{thm-main3}
Suppose that $\cQ$ is a collection of finite initially connected
initialized automata,  
and let $\cG$ denote the collection of simple groups
dividing the monoid of some automaton in $\cQ$. 
Then a Conway category satisfies the identities 
associated with the automata in $\cQ$ iff it satisfies 
the identities associated with the groups in $\cG$. 
\end{thm}

\begin{cor}
Suppose that $\cQ$ is a collection of finite automata,
or finite initially connected initialized automata. 
Then the Conway identities together with the identities 
associated with the (initialized) automata in $\cQ$ is complete for 
iteration categories iff for every finite simple group
$G$ there is an automaton $\bQ \in \cQ$ (or initialized automaton
$(\bQ,q) \in \cQ$) such that 
$G$ divides the monoid $M(\bQ)$.
\end{cor}

\begin{remark}
 This Corollary strengthens a result proved in 
\cite{EsIterationCategories} using some more sophisticated 
methods from the structure
theory of finite automata. Suppose that 
$\cQ$ is a collection of finite automata such that 
at least one nontrivial counter is a subautomaton 
of a restriction of an automaton in $\cQ$, i.e., such 
that there is an integer $n > 1$ and an automaton
$\bQ\in \cQ$ having an $n$-element subset of the states 
and an input letter $z$ inducing a cyclic
permutation of this subset. It was proved in 
\cite{EsIterationCategories} that the Conway idetities and 
the identities associated with the automata in a collection 
of finite automata $\cQ$  form a complete set of axioms
for iteration categories  iff for every finite simple group
$G$ there is an automaton $\bQ \in \cQ$ such that 
$G$ divides the monoid $M(\bQ)$.

We note that there is a class $\cQ$ of finite automata such that 
every finite group is isomorphic to the monoid 
of some automaton in $\cQ$ but there is no nontrivial counter 
that is isomorphic to a subautomaton of a restriction of any 
automaton in $\cQ$.
\end{remark}

For each $n \geq 3$, consider the $n$-state automaton 
$\bQ_n$ with state set $\{q_1,\ldots,q_n\}$ 
with an input letter inducing the cyclic permutation 
$(q_1\ldots q_n)$ and a 
letter inducing the transposition $(12)$, so that the monoid of $\bQ_n$ 
is the symmetric group $S_n$ of degree $n$. Then the Conway identities 
together with any 
infinite subsystem of the identities associated with the initialized automata 
$(\bQ_n,q_1)$ are complete for iteration categories
since every finite group may be embedded in any large enough symmetric group.
It is shown in 
\cite{EsIterationCategories} 
that the identity associated with $(\bQ_n,q_1)$ may be reduced to
\begin{eqnarray}
\label{eq_n}
(f\circ (\Delta_{A^2}\os \1{C})\circ 
\le f\circ \le \pi_1^{A \x C},\ (f\dr)^{n-2},\pi_2^{A \x C}\re, 
\ \pi_2^{A \x C}\re)\dr
&=&
(f\circ (\Delta_{A^2}\os \1{C}))\dr, 
\end{eqnarray}
where $f$ is any morphism $A^2 \x C \to A$. Thus, we obtain 
a simple proof of the fact that system consisting of 
the Conway identities and any infinite subset of the identities
(\ref{eq_n}) is  complete for iteration categories.

None of the complete systems consisting of the Conway identities and an 
infinite subcollection of the identities (\ref{eq_n}) 
is minimal, since each such system contains a complete proper subsystem.
 Similarly, no complete axiomatization of 
iteration theories consisting of the Conway identities 
and a subcollection of the automaton (or group) identities 
is minimal. 

{\bf Open problem} Does there exist a minimal complete set of identities
for iteration categories? 

Given a class $\cG$ of finite simple groups (including the trivial groups),
closed under division, there is a corresponding `variety' ${\bf V}_\cG$ 
of Conway categories axiomatized by the Conway identities and the identities 
$\Gamma(G)$ for $G \in \cG$. It contains the subvariety ${\bf V}_\cG^s$ consisting
of those categories in ${\bf V}_\cG$ with a single morphism $T \to A$
for any object $A$. (Here $T$ is the distinguished terminal object.)
These varieties are all different. 
A concrete description of the `free categories' (or rather,
free iteration theories) in these varieties 
was given in \cite{Espower}. It is possible that all 
varieties of Conway categories are of the form  ${\bf V}_\cG$ 
or  ${\bf V}_\cG^s$ . By the results 
of this paper, this holds iff every variety of Conway categories may be 
axiomatized by the Conway identities and a set of automaton identities,
and possibly the identity $f = g$, where $f,g: T \to A$.

{\bf Open problem} Describe the lattice of all varieties of Conway categories.

\section{Conclusion}

The equational properties of fixed point operations over cartesian
categories are captured by the axioms of iteration categories
which can be axiomatized by the Conway identities and
identities associated with (deterministic) finite automata,
or finite groups. We proved that in conjunction with the Conway
identities, the identity associated with a finite automaton 
implies each identity associated with every finite 
group that divides the monoid of the automton. We concluded that 
a system consisting of the Conway identities and a collection 
of the identities associated with the members $\bQ$ of a subclass 
$\cQ$ of finite automata is complete for iteration categories
iff for each finite simple group $G$ there is some automaton $\bQ \in \cQ$ 
such that $G$ divides the monoid of $\bQ$. This result establishes a connection between 
the equational theory of fixed point operations and the Krohn-Rhodes 
decomposition theorem for finite automata. Since several equational theories 
of computational interest have a finite axiomatization relatively 
to iteration categories, our result yields simple axiomatizations of 
these theories.

\end{document}